\documentclass[pre,twocolumn,showpacs,superscriptaddress]{revtex4-1}

\usepackage{graphicx}
\usepackage{amsmath}
\usepackage{amsfonts}
\usepackage{amssymb}
\usepackage{wasysym}
\usepackage{dsfont}
\usepackage{color,soul}
\usepackage{dcolumn}		
\usepackage{bm}			
\usepackage[pdftex, colorlinks, citecolor=blue, linkcolor=blue, urlcolor=blue]{hyperref}
\usepackage[all]{hypcap}		

\usepackage{amsmath}

\definecolor{redmarker}{rgb}{0.9,0.0,0.0}
\definecolor{greenmarker}{rgb}{0.0,0.6,0.0}
\definecolor{bluemarker}{rgb}{0.0,0.0,0.9}

\begin{document}
\title{Multiple binding sites for transcriptional repressors can \\ produce regular bursting and enhance noise suppression}

\author{Iv\'an M. Lengyel}
\affiliation{Instituto de Investigaci\'on en Biomedicina de Buenos Aires (IBioBA)\---CONICET\---Partner Institute of the Max Planck Society, Polo Cient\'{\i}fico Tecnol\'ogico, Godoy Cruz 2390, C1425FQD, Buenos Aires, Argentina}
\affiliation{Departamento de F\'{\i}sica, FCEyN UBA, Ciudad Universitaria, 1428 Buenos Aires, Argentina}%
\author{Luis G. Morelli}		\email{lmorelli@ibioba-mpsp-conicet.gov.ar}
\affiliation{Instituto de Investigaci\'on en Biomedicina de Buenos Aires (IBioBA)\---CONICET\---Partner Institute of the Max Planck Society, Polo Cient\'{\i}fico Tecnol\'ogico, Godoy Cruz 2390, C1425FQD, Buenos Aires, Argentina}
\affiliation{Departamento de F\'{\i}sica, FCEyN UBA, Ciudad Universitaria, 1428 Buenos Aires, Argentina}%
\affiliation{Max Planck Institute for Molecular Physiology, Department of Systemic Cell Biology, Otto-Hahn-Str.~11, D-44227 Dortmund, Germany}%

\date{\today}

\begin{abstract}
Cells may control fluctuations in protein levels by means of negative autoregulation, where transcription factors bind DNA sites to repress their own production.
Theoretical studies have assumed a single binding site for the repressor, while in most species it is found that multiple binding sites are arranged in clusters.
We study a stochastic description of negative autoregulation with multiple binding sites for the repressor.
We find that increasing the number of binding sites induces regular bursting of gene products.
By tuning the threshold for repression, we show that multiple binding sites can also suppress fluctuations.
Our results highlight possible roles for the presence of multiple binding sites of negative autoregulators.
\end{abstract}
\maketitle

\noindent
\section{Introduction} \label{sec_intro}
The state of living cells is determined by the molecules they produce and their numbers~\cite{AlbertsBook}.
Cells can control the production of molecules by means of gene regulatory networks~\cite{AlonBook}.
In these networks, transcription factors are key proteins that bind DNA to activate or repress synthesis~\cite{AlbertsBook,AlonBook}.
Negative autoregulation is a common component of these networks in which a transcription factor binds a specific site in the DNA and prevents its own synthesis~\cite{AlonBook}.
In \emph{E. coli}, about 40\% of transcription factors are negatively autoregulated~\cite{Thieffry1998, Rosenfeld2002, AlonBook}.

Since the number of some molecules in the cell can be small and the process of molecular synthesis is subject to fluctuations in the cellular environment, 
the resulting molecule numbers are in general noisy~\cite{Raser2005, Kaern2005, Sanchez2013r, Tsimring2014, Norman2015}.
It is thought that negative autoregulation can speed up response times~\cite{AlonBook, Rosenfeld2002} and tame fluctuations~\cite{Becskei2000, Dublanche2006, Kaern2005}.
Such circuits can suppress noise by regulating the number of molecules that a cell produces, 
permitting or repressing synthesis depending on the relative amount of the molecule~\cite{Becskei2000,Dublanche2006}.
However, negative autoregulation can also boost fluctuations if transcription factor binding is too strong~\cite{Stekel2008}.
A balance of timescales is key for negative autoregulation to be able to act as a noise suppressor~\cite{Singh2009, Gronlund2013}.
Furthermore, it has been argued that the cost of suppressing noise by negative autoregulation may be high due to fluctuations in intermediate signalling events~\cite{Lestas2010}.
Such intermediate events may also introduce effective delays and cause oscillations in molecule numbers~\cite{Morelli2007, Morant2009, Wang2014}.
%
%
%
\begin{figure}[b]
\centering
\includegraphics[width=\columnwidth]{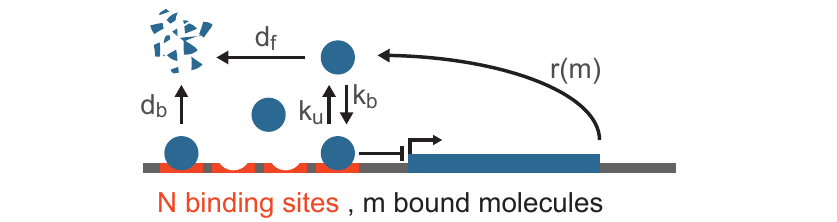}
\caption{ 
\emph{Negative autoregulation with multiple binding sites.}
Protein (blue circles) is synthesised from the information in the gene (blue stripe) at a rate $r(m)$. 
Protein binds to the $N$ binding sites (orange) at a rate $k_b$ and falls off at a rate $k_u$. 
Bound proteins repress production when more than $M$ proteins are bound (blunted arrow).
Free proteins are degraded at a rate $d_f$ and bound proteins at a rate $d_b$.
}
\label{f_sch}
\end{figure}

Despite its ubiquity, the effects of noise on negative autorregulation are still poorly understood.
Theoretical studies often assume a single binding site for a transcriptional repressor or an effective Hill function.
However, multiple binding sites for the same transcription factor are known to form clusters within regulatory domains~\cite{Takebayashi1994, Kazemian2013, Rydenfelt2014, Bothma2015}.
This feature has been widely observed in {\it E. coli}~\cite{Rydenfelt2014}, eukaryotes~\cite{Wagner1997, Wagner1999}, 
invertebrates~\cite{Lifanov2003, Berman2002} and vertebrates~\cite{Gotea2010}.
%
%
%
In fact more than half of the human genes contain clusters of binding sites for the same transcription factor~\cite{Gotea2010}.
This suggests that multiple binding sites for transcription factors may play an important role in gene regulation~\cite{Gotea2010, Thomas2005}.

The architecture of regulatory domains can affect bursting kinetics~\cite{To2010, Suter2011}.
In the presence of multiple binding sites, their occupation can have multiple states.
%
%
We hypothesize that stochastic switching will occur between these states and 
that this will set an additional timescale which affects fluctuations.
Here we study how fluctuations are affected by multiple binding sites, in the framework of a stochastic binding theory in which a single transcription factor represses its own production,~Fig.~\ref{f_sch}.

\section{Multiple stochastic binding} \label{sec_mult}
We consider a single gene encoding a protein that can bind to any of $N$ sites to repress its own synthesis.
We introduce negative autoregulation as a dependence of the production rate with the number $m$ of DNA bound proteins, $r=r(m)$, Fig.~\ref{f_sch}.
We focus on noncooperative binding, that is the affinity of proteins for binding sites $k_b$ is not affected by how many molecules are bound already.
Bound proteins fall off from the binding site with a rate $k_{u}$ per molecule.
%
Bound proteins have a decay rate $d_b$ and free proteins have a decay rate $d_f$ per molecule.

The statistics of the stochastic process can be described by the probability distribution $P(n,m,t)$ to have $n$ free proteins and $m$ bound proteins at time $t$~\cite{VanKampen1992}.
This probability distribution obeys the master equation
\begin{equation} \label{master_eq}
\begin{aligned}
& \frac{d P(n,m,t)}{dt}= r(m) \left( P(n-1,m,t)-P(n,m,t) \right) \\
&+ k_b \left( (N-(m-1))(n+1) P(n+1,m-1,t) \right) \\
&- k_b \left(  (N-m)n P(n,m,t) \right) \\
&+ k_u \left(  (m+1) P(n-1,m+1,t)  -   m P(n,m,t) \right) \\
&+ d_f \left( (n+1) P(n+1,m,t) - n P(n,m,t) \right)  \\
&+ d_b \left( (m+1)P(n,m+1,t)-mP(n,m,t) \right) \, .
\end{aligned}
\end{equation}
The first line accounts for molecule production, 
the second and third for the binding process, 
the fourth for the unbinding process, 
and the two last lines for degradation of free and bound molecules.

The synthesis rate $r(m)$ can be adjusted to describe different regulatory mechanisms.
For simplicity, here we consider the case of a sharp threshold:
synthesis occurs at a rate $r_0$ while there are $M$ or less bound proteins, 
and it is fully repressed when there are more than $M$ bound proteins
\begin{equation} \label{eq_rm}
r(m)=
\begin{cases}
r_0 &\text{if  }  0\leq m \leq M \, , \\  
0    &\text{if  }  M+1\leq m \leq N \, .
\end{cases}
\end{equation}
Although a function $r(m)$ has not been determined experimentally,
elegant experiments have determined the regulatory relationship 
between transcription factor concentration and gene activity~\cite{Xu2015}.
The form of the synthesis rate as a function of the number of bound transcription factors 
that we chose here gives rise to effective regulatory functions which are consistent 
with these experimental observations~\cite{Lengyel2014}.
Additionally, below in Section~\ref{sec_noise} we discuss the case of a graded synthesis rate 
to show that our conclusions do not depend strongly on this choice.

In general, changing regulatory domain architecture will affect the total number of proteins mean value and fluctuations.
Here we are interested in how changes to regulatory domains architecture affect fluctuations, under the assumption that autoregulation controls the level of proteins to keep it at some constant functional value in steady state~\cite{AlonBook}.
With similar mean expression levels, noise can vary by more than one order of magnitude when the regulatory domain architecture is changed~\cite{Sharon2014}.
Because adding binding sites to the regulatory domain may change the mean number of molecules $\langle n_T \rangle = \langle n+m \rangle$, to keep it constant we adjust the synthesis rate $r_0$.
In the following we choose the value $\langle n_T \rangle = 20$ for illustration~\cite{Stekel2008,Grima2012,Gronlund2013}. 

We generate trajectories satisfying equation (\ref{master_eq}) by means of a standard Gillespie algorithm~\cite{Gillespie1976, Gillespie2007}.
We obtain the free $n(t)$, bound $m(t)$ and total $n_{T}(t)=n(t)+m(t)$ number of proteins.
The total number of proteins $n_T$ is useful to compare theoretical results with experiments which do not distinguish free and bound molecules.
%
%
\begin{figure}[t]
\centering
\includegraphics[width=\columnwidth]{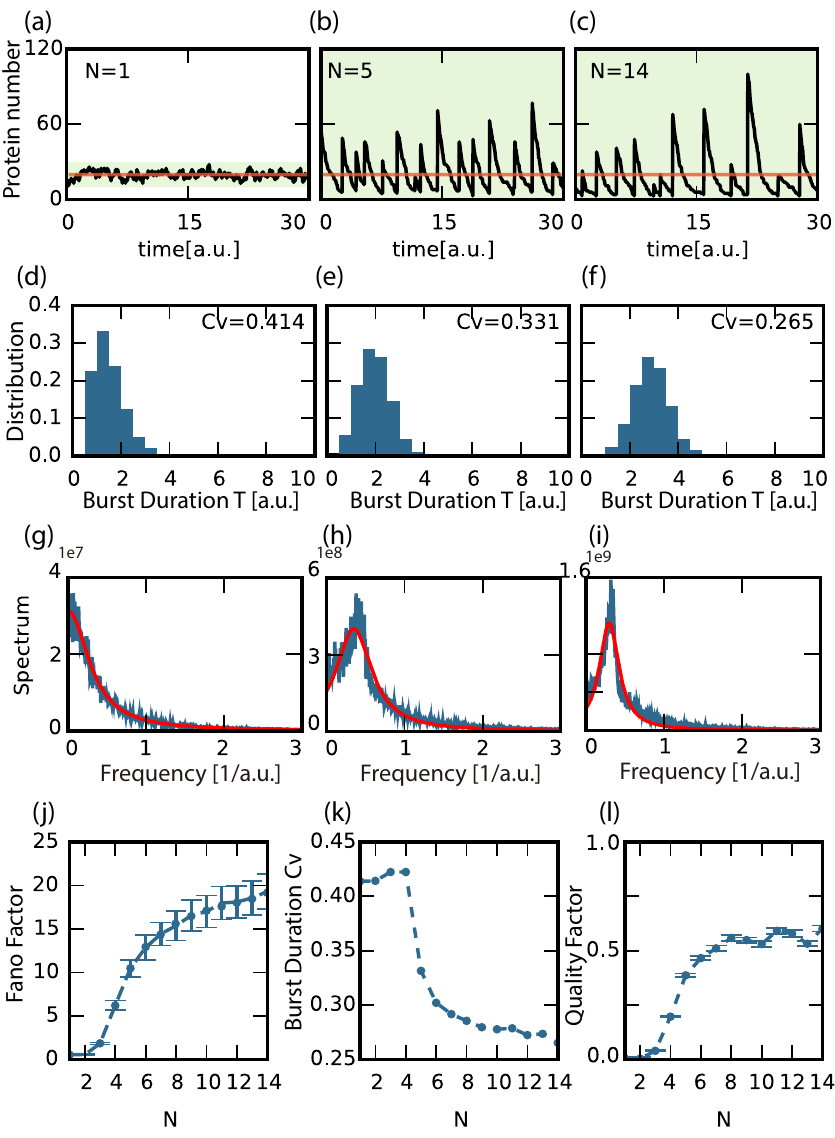}
\caption{
\emph{Increasing the number of binding sites $N$ increases size and temporal precision of fluctuations.}
Time series of the total number of proteins $n_T(t)$ from stochastic simulations for $M=0$ and (a) $N=1$, (b) $N=5$ and (c) $N=14$.
Orange line indicates the mean value $\langle n_T \rangle$ and the shaded region indicates the variance bounds. 
(d-f) Burst duration distribution for the panels (a-c) resulting from $N_s=100$ stochastic simulations of 200 [a.u.] for each parameter set
(g-i) Power spectral density (blue line) from data in panels (a-c) resulting from averaging 100 times series of length 200 [a.u.]. 
%
%
Fit to the spectrum with a Lorentz Distribution (red line) to obtain the central frequency $f_c$ and the half width $\Delta_f$.
(j) Fano factor $F$ of the total number of proteins computed from times series. 
(k) Coefficient of variation of burst duration $T$ as a function of the number of binding sites $N$.
(l) Quality factor $Q$ obtained from the power spectral density of time series.
Parameters: $k_b=k_u=335.5$, $d_f=d_b=1$;
the value of $r_0$ is adjusted to keep $\langle n_T \rangle$ constant: (a) $r_0=409$, (b) $r_0=1743696$ and (c) $r_0=9598311$.
(j,k) Dots correspond to the mean value and error bars to the standard deviation over $N_s=100$ stochastic simulations.
(l) Error bars are the error of the fit. 
(j-l) Dots are joined by a dashed line to guide the eye.}
\label{f_osc}
\end{figure}

\section{Regular bursting}\label{sec_regular}
We first study the case $M=0$ in which a single bound protein fully represses production, and we analyse the effects of varying number of binding sites $N$.
%
We consider the case in which bound proteins are degraded at the same rate as free proteins, $d_f=d_b=1$.
Deviations of the total number of proteins $n_T$ from its mean value $\langle n_T \rangle$ increase with $N$, Fig.~\ref{f_osc}(a-c).
Adding binding sites to the system leads to bursty kinetics, with increasing burst size and consequently longer decay times.
This is reflected on an increasing Fano factor, defined as the variance to mean ratio $F = \sigma_{n_T}^2 / \langle n_T \rangle$, Fig.~\ref{f_osc}(j).

These bursts occur with some temporal regularity.
The time interval between two consecutive maxima naturally defines burst duration $T$. 
Varying $N$ we compute the burst duration distribution from stochastic simulations, Fig.~\ref{f_osc}(d-f).
%
%
Burst duration becomes longer while the relative dispersion of the distribution around the mean decreases with increasing $N$,
as reflected in the coefficient of variation $CV=\sigma_T / \langle T \rangle$, Fig.~\ref{f_osc}(k).
Thus, the characteristic time governing burst timing becomes both longer and more precise.

Temporal precision can be further characterized by the quality factor 
\begin{equation} \label{eq_q}
Q = \frac{f_0} {\pi \Delta f} \, .
\end{equation}
where $\Delta f$ is the bandwidth and $f_0$ the main frequency.
Higher values of $Q$ indicate a narrower frequency spectrum and a more defined characteristic time~\cite{Morelli2007}.
%
%
To estimate the quality factor we fit a Lorentzian distribution to the power spectral density, see Appendix~\ref{sec_hilbert} for the details.
From this fit, Fig.~\ref{f_osc}(g-i), we determine the main frequency $f_0$ and half power bandwidth $\Delta f$.
The quality factor increases with $N$ reflecting the observed increased regularity of burst duration, Fig.~\ref{f_osc}(l).  
This indicates an increase in bursting temporal precision, although the values of the quality factor remain low.

As indicated in Section \ref{sec_mult} we keep the mean value of the total number of proteins $\langle n_T \rangle$ fixed by adjusting the value of the synthesis rate $r_0$.
The large values of $r_0$ employed in Fig.~\ref{f_osc} are required to balance the large values of binding affinity that occur as $N$ is increased, see Appendix~\ref{sec_prop}.

The multiple binding sites act as a buffer for the bound proteins:
as long as the occupation number $m$ is larger than one, there can be fluctuations in $m$ and the system is still repressed.
In this way, multiple binding sites give rise to an effective timescale in the system, which manifests as the time taken to exponentially decay from the maximum to the minimum value of total number of proteins.
In Section \ref{sec_bound} below we further discuss the mechanism underlying this observation.
\begin{figure}
\centering
\includegraphics[width=\columnwidth]{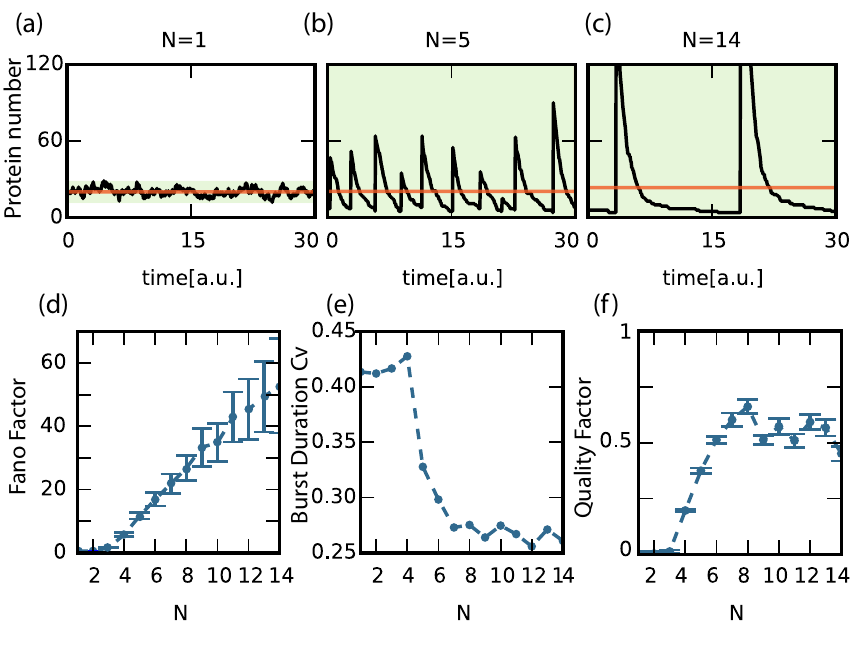}
\caption{
\emph{Increasing the number of binding sites $N$ increases size and temporal precision of fluctuations when bound proteins are not degraded.}
Time series of the total number of proteins $n_T(t)$ from stochastic simulations for $M=0$ and (a) $N=1$, (b) $N=5$ and (c) $N=14$.
Orange line indicates the mean value $\langle n_T \rangle$ and the shaded region indicates the variance bounds. 
(d) Fano factor $F$ of the total number of proteins computed from times series (blue line with errors).
(e) Coefficient of variation of burst duration $T$ as a function of the number of binding sites $N$.
(f) Quality factor $Q$ obtained from the power spectral density of time series.
Parameters: $k_b=k_u=335.5$, $d_f=d_b=1$;
note the value of $r_0$ is adjusted to keep $\langle n_T \rangle$ constant: (a) $r_0=390$, (b) $r_0=2257059$ and (c) $r_0=2257059$.
(d-f) each data point is computed from 100 time series of 200 [a.u.] long.
(d,e) Dots correspond to the mean value and error bars to the standard deviation over $N_s=100$ stochastic simulations.
(f) Error bars are the error of the fit. 
(d-f) Dots are joined by a dashed line to guide the eye.}
\label{f_osc_nodeg}
\end{figure}

\subsection*{Degradation of bound proteins is not required for regular bursting}
In eukaryotic cells some proteins are actively degraded by the complex machinery of the proteasome~\cite{AlbertsBook}.
The molecular machines of the proteasome recognise specific domains in the proteins surface and target these proteins for destruction.
When a transcription factor binds DNA, these protein domains may or may not become obstructed. 
%
Therefore, bound and free proteins can have different degradation rates.
So far we have addressed the case in which both bound and free proteins are degraded at the same rate $d_f=d_b=1$.
Here we study the case in which bound proteins cannot be degraded, setting $d_b=0$.

Adding $N$ binding sites with fixed $M=0$ also leads to bursty kinetics, Fig.~\ref{f_osc_nodeg}(a-c).
Since there is no degradation of bound proteins, the number of bound proteins decrease only when there is an unbinding event.
Thus, the probability of finding $m=0$ is lower leading to longer decaying times under the mean and higher burst maximums. 
The fluctuations around the mean quantified by the Fano factor increase with the number of binding sites $N$, Fig.~\ref{f_osc_nodeg}(d).
The bursty fluctuations have a characteristic time that becomes longer as $N$ increases. 
The dispersion around the mean of the maximum distribution decreases indicating that the characteristic time becomes more precise although the decay times are longer, Fig.~\ref{f_osc_nodeg}(e).
This increased regularity is also reflected in the quality factor $Q$, Fig.~\ref{f_osc_nodeg}(f).
%

\section{Noise suppression}\label{sec_noise}
We now turn to the effects of changing the threshold $M$ for a fixed value of total number of binding sites $N$.
We first consider the case in which bound proteins are degraded at the same rate as free proteins, $d_f=d_b=1$.
As $M$ changes, we adjust the synthesis rate $r_0$ accordingly to keep the mean value $\langle n_{T} \rangle$ fixed.

\begin{figure}[t]
\centering
\includegraphics[width=\columnwidth]{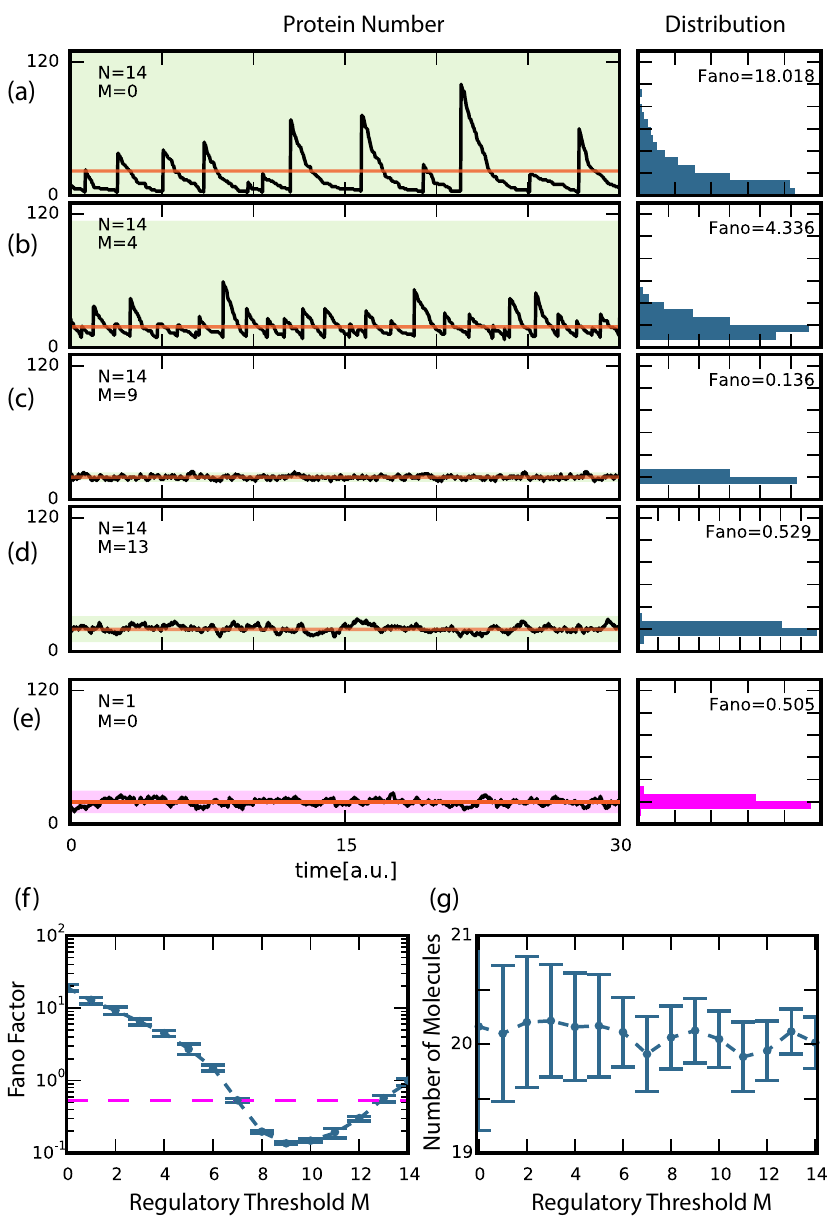}
\caption{
\emph{Multiple binding sites can reduce fluctuations without increasing the turnover.} 
(a-e) Time series of the total number of proteins (left panels) and corresponding histograms (right panels).
(a-d) $N=14$ and $M=0$, 4, 9 and 13, and (e) single binding site autoregulation $N=1$, $M=0$.
Orange line indicates the mean value and shaded region the variance bounds. 
(f) Fano factor and (g) number of molecules degraded per unit of time, obtained from time series. 
The single binding site result is displayed for reference (dashed magenta line).
 Parameters: $k_b=k_u=335.5$, $d_f=d_b=1$.
In (f) and (g) we first computed the Fano factor/Number of molecules from time series of 200 [a.u.] long, and then averaged over 100 realisations.
Dots correspond to the mean value and error bars to the standard deviation.
Dots are joined by a dashed line to guide the eye.
}
\label{f_noi}
\end{figure}
Increasing the threshold $M$, fluctuations decrease to a minimum and then rise again, Fig.~\ref{f_noi}(a-d). 
This minimum occurs when synthesis 
events are driven by stochastic binding and unbinding of proteins, see Section~\ref{sec_bound} for a further discussion about the mechanism. 
For lower values of $M$, fluctuations around the mean value of bound proteins $\langle m \rangle$ are not sufficient to give rise to synthesis events because a small number of molecules is enough to overcome the low threshold.
For higher values of $M$, fluctuations that bring $m$ below threshold are more frequent and synthesis rate is slower, resulting in a lax control of the mean value of proteins $\langle n_T \rangle$.

Noise suppression with multiple binding sites outperforms the single binding site control system, see its smaller variance and Fano factor, Fig.~\ref{f_noi}(c,e).
This stronger noise suppression occurs for a range of $M$ values, Fig.~\ref{f_noi}(f).
Even in cases in which single binding site regulation $N=1$ increases fluctuations~\cite{Stekel2008, Gronlund2013, Huang2014}, the presence of multiple binding sites may bring noise below the level for unregulated synthesis.
Moreover, the turnover of molecules does not change with the threshold value, that is the number of molecules degraded per unit time is constant for different values of $M$, Fig.~\ref{f_noi}(g).
This indicates that this noise suppression mechanism comes at no extra cost for the system.

\subsection*{Bound protein degradation is not required for noise suppression}
\begin{figure}[t]
\centering
\includegraphics[width=\columnwidth]{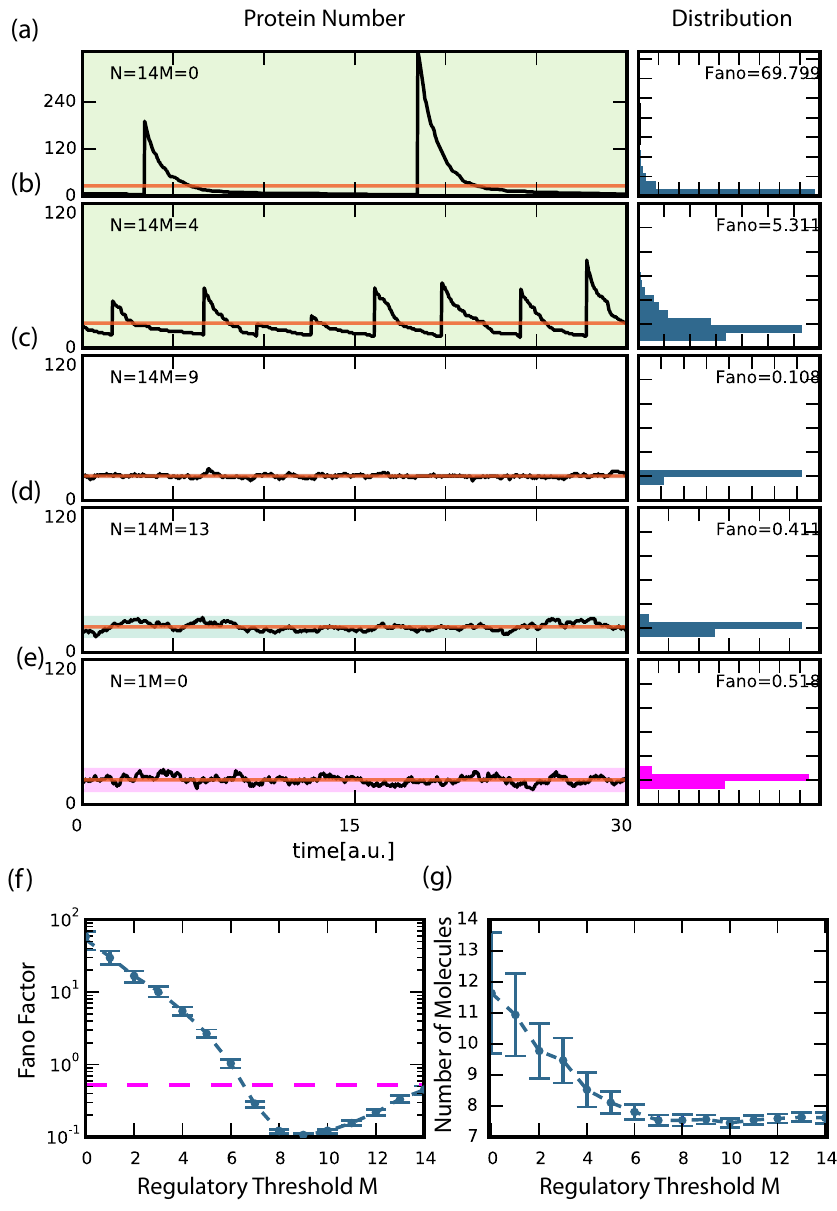}
\caption{
\emph{Multiple binding sites can reduce fluctuations without increasing the turnover when bound proteins are not degraded. }
(a-e) Time series of the total number of proteins (left panels) and corresponding histograms (right panels).
(a-d) $N=14$ and $M=0$, 4, 9 and 13, and (e) single binding site autoregulation $N=1$, $M=0$.
Orange line indicates the mean value and shaded region the variance bounds. 
(f) Fano factor and (g) number of molecules degraded per unit of time, obtained from time series (blue line) for $N=14$. 
The single binding site result is displayed for reference (dashed magenta line).
Parameters: $k_b=k_u=335.5$, $d_f=1$, $d_b=0$.
In (f) and (g) we first computed the Fano factor/Number of molecules from time series of 200 [a.u.] long, and then averaged over 100 realisations.
Dots correspond to the mean value and error bars to the standard deviation.
}
\label{f_noise_nodeg}
\end{figure}

%
%
So far we analysed noise suppression in a situation where bound and free proteins are degraded at the same rate.
Next we analyse the case in which bound proteins are not degraded, that is $d_b=0$.
For fixed $N$, noise suppression can be achieved by changing the value of the regulatory threshold $M$, Fig.~\ref{f_noise_nodeg}(a-d).
There is a value which optimise the noise for fixed $N$, and there is a range of values for which the noise suppression is larger than for the single site case, Fig.~\ref{f_noise_nodeg}(e,f).
Turnover decreases as the regulatory threshold $M$ increases due to a sequestration effect, since bound proteins are not degraded, Fig.~\ref{f_noise_nodeg}(g).
%
%

\subsection*{A sharp threshold synthesis rate is not required for noise suppression}
%
%
%
%
In previous Sections we analysed the case of a synthesis rate $r(m)$ with a sharp regulatory threshold, Eq.~(\ref{eq_rm}). 
Here we analyse the case of a graded synthesis rate which decreases linearly with the number of bound proteins 
%
%
%
\begin{equation} \label{ec_grad}
r(m)=
\begin{cases}
r_0 (1 - m/ (M+1))	&\text{if  }  0\leq m \leq M \\
0				&\text{if  }  M+1\leq m \leq N \, .
\end{cases}
\end{equation}
The maximum value of bound proteins for which the system can synthesise new molecules is $m=M$, if $m > M$ the system is fully repressed~\footnote{
For $M=0$ this graded synthesis rate is equivalent to a sharp threshold~Eq.~(\ref{ec_grad}).
For this reason we do not discuss regular bursting with a graded synthesis rate in Section~\ref{sec_regular}.
However with multiple binding sites $N>1$ and $M > 0$ Eq.~(\ref{ec_grad}) is different from Eq.~(\ref{eq_rm}).}.
%

\begin{figure} 
\centering
\includegraphics[width=\columnwidth]{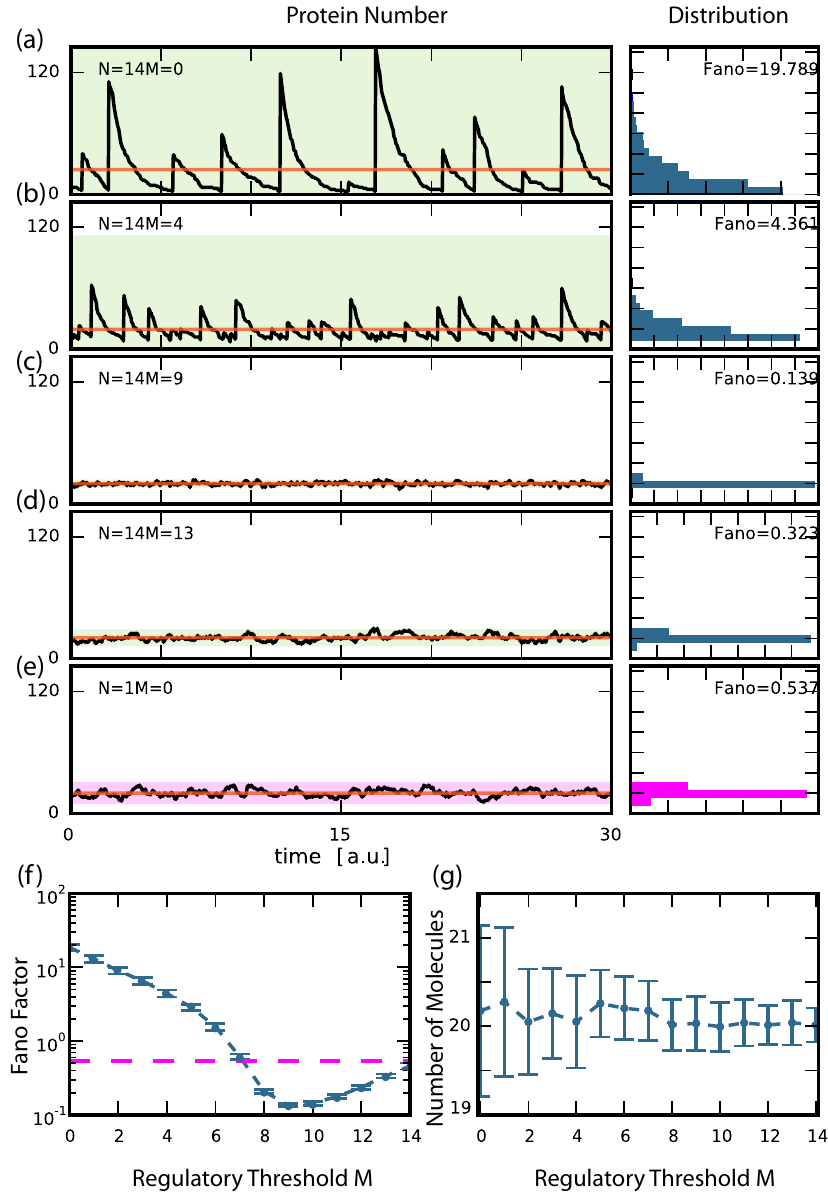}
\caption{
\emph{Multiple binding sites can reduce fluctuations without increasing the turnover in the case of a graded synthesis rate.}
(a-e) Time series of the total number of proteins (left panels) and corresponding histograms (right panels).
(a-d) $N=14$ and $M=0$, 4, 9 and 13, and (e) single binding site autoregulation $N=1$, $M=0$.
Orange line indicates the mean value and shaded region the variance bounds. 
(f) Fano factor and (g) number of molecules degraded per unit of time, obtained from time series (blue line) for $N=14$.  
The single binding site result is displayed for reference (dashed magenta line).
 Parameters: $k_b=k_u=335.5$, $d_f=d_b=1$.
%
%
In (f) and (g) we first computed the Fano factor/Number of molecules from time series of 200 [a.u.] long, and then averaged over 100 realisations.
Dots correspond to the mean value and error bars to the standard deviation.
}
\label{f_noise_grad}
\end{figure}

Following the same approach as in the sharp regulatory threshold case, we analyse noise levels when changing $M$ for a fixed value of $N$. 
We analyse the case in which bound and free proteins have the same degradation rate $d_f=d_b=1$.
As elsewhere, we adjust the value of $r_0$ to keep the mean value $\langle n_T \rangle$ fixed for different values of $M$. 
As $M$ increases the noise decreases and then increase again Fig.~\ref{f_noise_grad}(a-d). 
The noise suppression mechanism with a graded synthesis rate can enhance noise suppression and 
outperforms the single site case $N=1$ and $M=0$, 
Fig.~\ref{f_noise_grad}(e), for a range of values of $M$, Fig.~\ref{f_noise_grad}(f). 
The turnover does not change while $M$ changes, Fig.~\ref{f_noise_grad}(g).
Thus, a sharp threshold is not required for noise suppression by multiple binding sites.
%
%

\section{Regular bursting and noise suppression are robust}\label{sec_robust}
\begin{figure}
\centering
\includegraphics[width=\columnwidth]{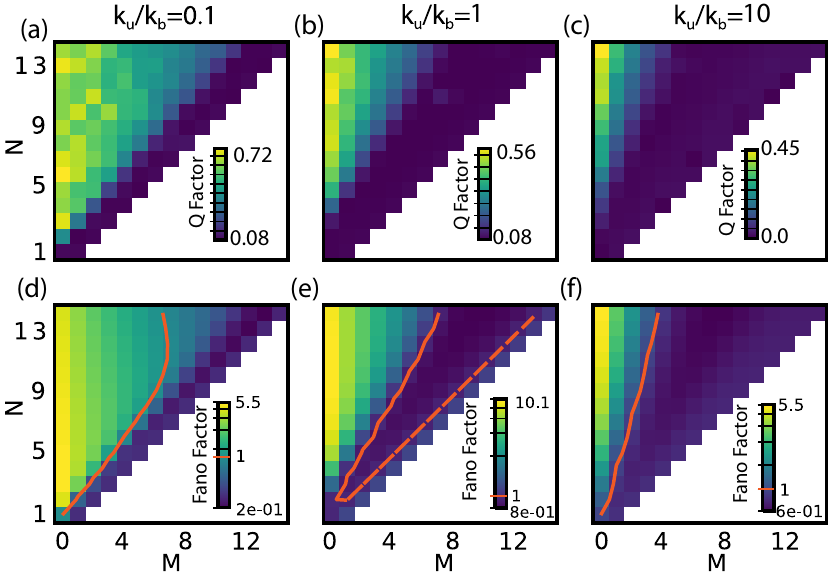}
\caption{
\emph{Regular bursting and noise suppression are robust to different binding / unbinding ratios.}
(a-c) Quality factor maps for different numbers of binding sites $N$ and threshold $M$.
(d-f) Relative Fano factor $F_R$.
Orange lines correspond to $F_R=1$.
Parameters: (a) $k_u=1$, $k_b=10$, (b) $k_u=k_b=335$ and (c) $k_u=10, k_b=1$.
In all panels $d_f=d_b=1$.
Mean value of quality and Fano factor are averaged over $N_s=100$ stochastic simulations.}
\label{f_squ}
\end{figure}

So far we focused on a set of parameters in which the binding and unbinding rates are equal, $k_b=k_u$.
%
%
We now lift this constraint and analyse the effects of multiple binding sites and changing the threshold $M$ for different $k_b$ and $k_u$ ratios, Fig.~\ref{f_squ}.
In the three different cases spanning three orders of magnitude, adding multiple binding sites generates regular bursting for low $M$, Fig.~\ref{f_squ}(a-c).
The smaller the ratio $k_u/k_b$, the larger the quality factor and the region in $\{N,M\}$ space where regular bursting is enhanced.
Increasing the binding affinity generates longer decay times of bursts together with a larger memory effect. 

The noise suppression mechanism discussed above is present for a wide range of $k_u/k_b$ ratios, Fig.~\ref{f_squ}(d-f).
We introduce the relative Fano factor $F_R$, defined as the Fano factor normalised to its single binding site value ($N=1$ and $M=0$).
Values of $F_R$ smaller than one reveal a region in $\{N,M\}$ space in which noise suppression outperforms the single binding site case, Fig.~\ref{f_squ}(d-f).
For any number of binding sites $N$ there is a threshold $M$ that reduces fluctuations.
As the ratio $k_u/k_b$ increases the region of noise suppression becomes larger and the relative suppression effect stronger.
%
%
This is in line with the finding that weak binding is critical for the fidelity of autoregulation~\cite{Gronlund2013}.

The particular form of the synthesis rate $r(m)$ Eq.~(\ref{eq_rm}) is not crucial for regular bursting and noise suppression. 
Both phenomena are observed also with a graded synthesis rate that decreases linearly with $m$ (Sections~\ref{sec_regular} and \ref{sec_noise}, and Fig.~\ref{f_noise_grad}).
While here we have chosen the same degradation rate for free and bound proteins,
Regular bursting and noise suppression also occur in a system in which bound proteins cannot be degraded (Section~\ref{sec_regular} and \ref{sec_noise} and Figs.~\ref{f_osc_nodeg} and \ref{f_noise_nodeg}).
These observations suggest that the reported phenomena are robust and do not depend on our simplifying assumptions.
\begin{figure} 
\centering
\includegraphics[width=\columnwidth]{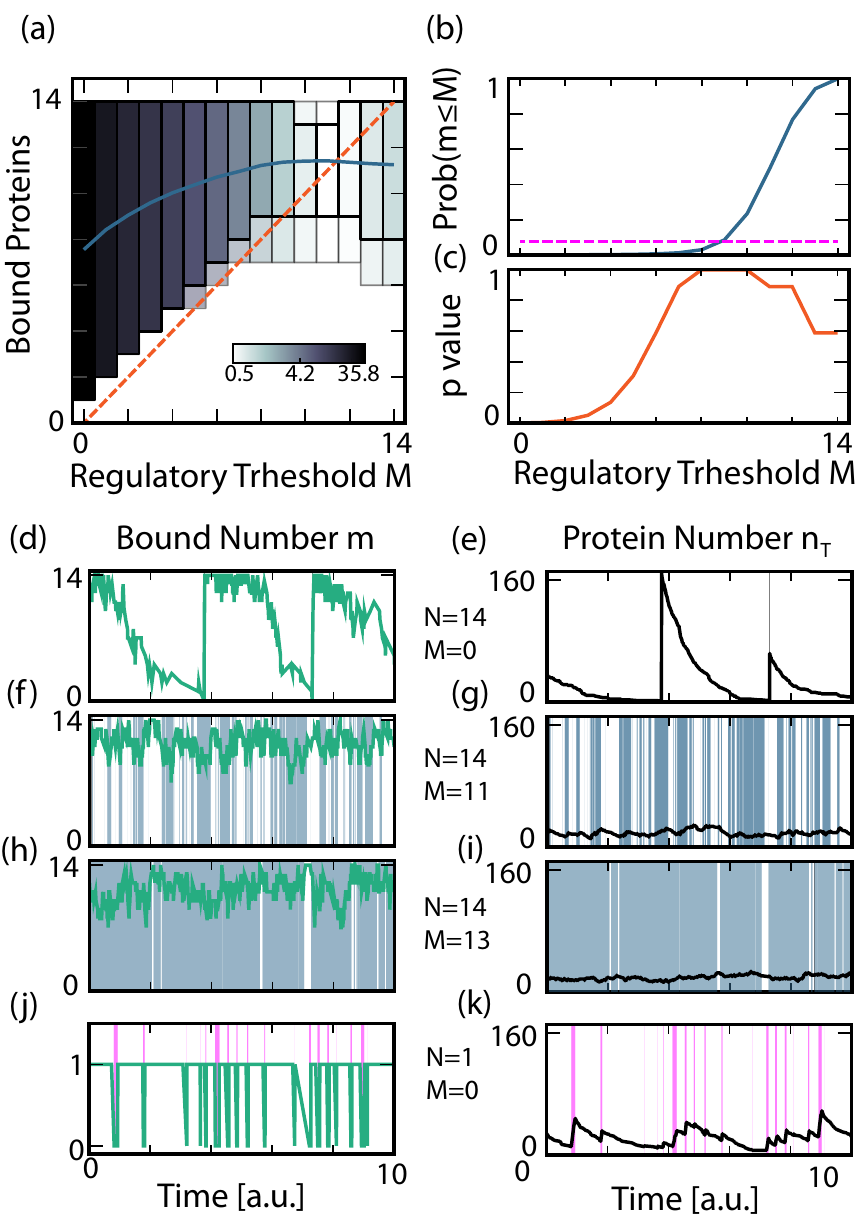}
\caption{\emph{Bound protein dynamics.}
(a) Mean value and confidence interval of bound protein $m$.
Solid blue line indicates the mean value of $m$ for the different regulatory thresholds $M$.
Dashed orange line indicates the reference value of the regulatory threshold $M$.
Shaded boxes indicate 90\% (dark) and 99\% (light) confidence intervals. 
The color scale goes from highest (black) to lowest (white) Fano factor.
(b) Probability $\mbox{Prob}(m\leq M)$ of finding synthesis unrepressed for the case $N=14$, solid blue line.
%
%
Dashed magenta line indicates a reference value of the probability for $N=1$ and $M=0$.
(c) p-value of two sample Kolmogorov-Smirnov test comparing the distribution of bound proteins in two situations, 
with ($r, d_f, d_b \neq 0$) and without synthesis and degradation ($r = d_f =d_m = 0$) .
(d, f, h, j) Time series of the bound number of proteins $m(t)$.
(e, g, i, k) Time series of total number of proteins $n_T(t)$.
In panels (d-k) shaded region (grey or magenta) indicates when $m(t) \leq M$.
Parameters: $k_b=k_u=1$, $d_f=d_b=1$.
(a-c) Statistics computed over 100 realisations of 200 [a.u.] for every $M$.
}
\label{f_sup}
\end{figure}

\section{Bound protein dynamics drives regular bursting and noise suppression}  \label{sec_bound}
This Section discusses in more detail the mechanism behind regular bursting and noise suppression reported in the previous Sections.
We showed that adding multiple binding sites gives rise to regular bursting, Figs.~\ref{f_osc} and~\ref{f_osc_nodeg}.
We argued that this occurs because multiple binding sites introduce an effective timescale in the system and act as a buffer of bound proteins.
We further showed that fluctuations are minimised for some value of the threshold $M$, Figs.~\ref{f_noi}, \ref{f_noise_grad} and~\ref{f_noise_nodeg}.
We argued that the minimum occurs when synthesis events are short and frequent
because the dynamics of the bound proteins $m$ is dominated by the processes of stochastic binding and falling off.
Below we expand on these ideas and discuss how regular bursting and noise suppression emerge as a consequence of the dynamics of bound proteins in the presence of multiple binding sites.

With multiple binding sites and low values of $M$, bound protein $m$ fluctuations are not enough to cross the threshold, Fig.~\ref{f_sup}(a) and (d).
No production events are possible while $m>M$, so the system is inhibited most of the time, Fig.~\ref{f_sup}(b).
Only after degradation of most proteins the number of proteins becomes very small and $m\leq M$, making synthesis events possible for a brief period of time, Fig.~\ref{f_sup}(d) and (e).
During this short interval of time a burst of synthesis occurs overshooting the total number of proteins $n_T$ over the mean, Fig.~\ref{f_noi}(a) and (b) and Fig.~\ref{f_sup}(e).
This overshoot over the mean together with long degradation times lead to large fluctuations in the number of proteins and high noise. 
These excursions over the mean have a characteristic timescale set by the time that takes the bound proteins to go below $M$ after a synthesis burst, Fig.~\ref{f_sup}(d).
Stochastic fluctuations of $m$ which do not cross the threshold $M$ do not lead to synthesis events.
In this way multiple binding sites confers a buffer of bound proteins that avoids synthesis events due to stochastic fluctuations.

For high values of $M$ bound protein $m$ fluctuations are enough to cross the threshold and the probability of finding the system active is large, Fig.~\ref{f_sup}(a) and (b).
Furthermore, for the highest values of $M$ the mean value is below threshold $\langle m \rangle \leq M$, Fig.~\ref{f_sup}(a) and (h), indicating that most of the time the system can produce new molecules.
In this situation the probability of finding the gene active is large and the synthesis rate $r$ is small in order to keep $\langle n_T \rangle =20 $, Fig.~\ref{f_sup}(i).
On one hand, this small synthesis rate leads to a slow response that keeps the total number of proteins below the mean.
On the other hand, since the number of free binding sites is typically low in this situation, binding events that overcome the threshold and inhibit the synthesis are not frequent, Fig.~\ref{f_sup}(h).
This small number of free sites and lack of inhibition keeps synthesis going even when the number of proteins is above the mean value $n_t\geq \langle n_t \rangle $.
The slow response in both situations results in a lax control which increases deviations from the mean. 
This causes poor noise regulation for high values of regulatory threshold $M$.

When the mean value of bound proteins is close to the threshold noise suppression is optimal, Fig.~\ref{f_sup}(a) and (f).
Fluctuations around the mean value are enough to cross the threshold frequently and the system is able to synthesise new molecules.
These excursions of the bound proteins are mainly due to stochastic binding and unbinding events.
To show this, it is instructive to consider the particular case where there is no degradation and no production, with fixed $n_T = \langle n_T \rangle $.
This particular case provides a reference distribution, where the dynamics of $m(t)$ is only due to stochastic binding and unbinding events.
We can compare this reference distribution with the distribution of $m(t)$ for different values of thresholds $M$ and $N=14$ in the general case.
We compare the distributions for different $M$ with the particular case with a two sample Kolmogorov-Sminorv test.
We test whether the values of $m$ of the different cases and the reference case come from the same distribution.
High p-values indicates that we cannot reject the hypothesis that the distributions of the two samples are the same.
Where the noise suppression is minimum we see that the p-values are higher, Fig.~\ref{f_sup}(c).
This indicates that excursions of the bound proteins above and below threshold are mainly due to stochastic binding and unbinding, and not because of changes in the total number of proteins due to synthesis and degradation of molecules.
Thus, since the mean value and the threshold are close, the excursions and the short and frequent synthesis events are due to fluctuations in the bound proteins.

We include the one-binding-site case $N=1$, $M=0$, Fig.~\ref{f_sup}(b,j) and (k) for reference, to highlight the effects of adding multiple binding sites.
Synthesis events occur in bursts, and fluctuations from the mean value are large since synthesis events are rare.
%
%
However there is no characteristic timescale in the system since fluctuations in the bound proteins drive the system in and out from synthesis and there is no buffer, Fig.~\ref{f_sup}(j).

\subsection*{Binding and unbinding kinetics.}
%
%
%
%
We have discussed how the regulatory domain architecture specified by 
the number of binding sites $N$ and the regulatory threshold $M$
controls the mechanism giving rise to regular bursting and noise suppression.
%
%
Here we show how binding kinetics affects this mechanism by varying the rates $k_b$ and $k_u$ for a fixed regulatory architecture. 
\begin{figure}
\centering
\includegraphics[width=\columnwidth]{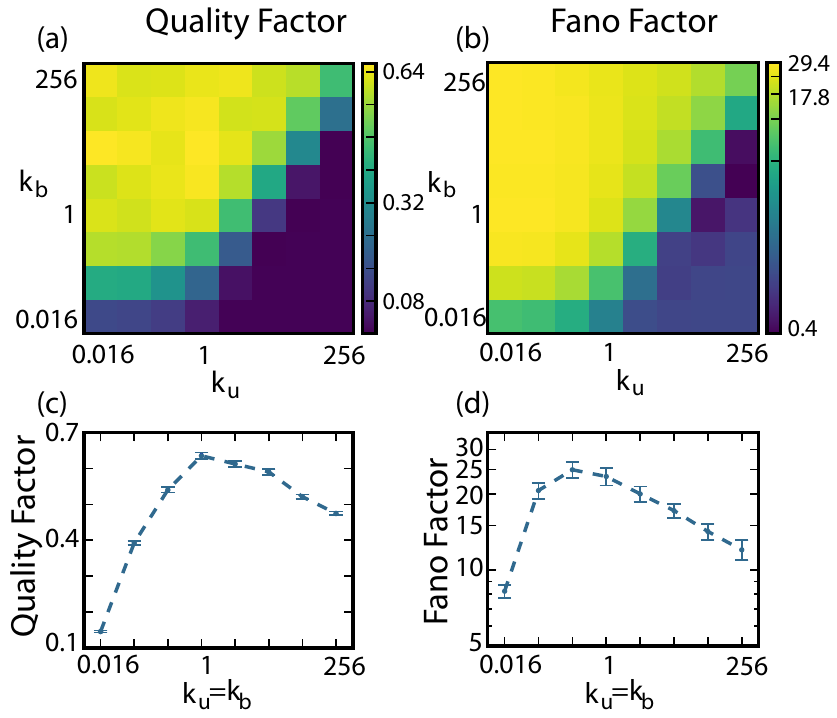}
\caption{
\emph{The ratio  $k_u/k_b$ controls noise and bursting temporal precision.}
(a) Quality factor and (b) Fano factor maps for different values of binding rate $k_b$ and unbinding rate $k_u$.
(c) Quality factor and (d) Fano factor for different rates with $k_u=k_b$.
Parameters: $N=11$, $M=1$,$d_f=d_b=1$.
(a,c) Data points correspond to the quality factor obtained from the power spectral density of 100 realisations 200 [a.u.] long, and error bar is the error in the quality factor computed from the least squares fit.
(b,d) Data points correspond to the mean Fano factor of 100 realisations of 200 [a.u.] long, and error bar is the standard deviation.
}
\label{f_kbku_Qn}
\end{figure}

Small $k_u/k_b$ ratios increase the Fano and the quality factor, Fig.~\ref{f_kbku_Qn}(a,b).
The increase in the Fano factor indicates that there is more dispersion in the total number of proteins around the mean value.
The increase in the quality factor indicates that this dispersion comes with increased temporal precision and a characteristic timescale.
As the ratio $k_u/k_b$ decreases, binding becomes more stable. 
%
%
A stronger binding than unbinding rate generates more stably bound proteins and larger decay times, setting a characteristic timescale or increasing its length.

At large $k_u/k_b$ ratios binding is unstable, Fig.~\ref{f_kbku_Qn}(a,b). 
In this scenario there are large fluctuations in the number of bound proteins under the threshold $M$.
Production events are frequent, but the number of molecules produced in each event is small.
This leads to small fluctuations around the mean value of total number of proteins.
Since production events are driven mainly by stochastic fluctuations of $m$ below $M$, there is no typical characteristic time in the system.
Thus, a stronger unbinding than binding rate generates small fluctuations without a characteristic timescale and with a smaller Fano factor.

For $k_b=k_u=k$, the Fano and quality factor exhibit a non-monotonic behaviour as a function of $k$, Fig.~\ref{f_kbku_Qn}(c,d). 
For low values of $k$ the binding probability is low.
Binding events are not so frequent and $\langle m \rangle$ is small.
Since $\langle m \rangle$ is small and $M=1$, fluctuations of bound proteins that cross the threshold are frequent and there is no characteristic timescale in the system.
When we increase the value of $k$ the probability of binding increases.
The mean value of bound proteins $\langle m \rangle$ is not close to the threshold $M$ as before.
This give rise to a memory effect in the bound proteins and a characteristic timescale in the system.
For even larger values of $k$ both fluctuations and quality factor decrease.
The probability of binding and unbinding is large, so binding and unbinding events are frequent.
This generates high frequency fluctuations in the number of bound proteins.
As the burst decays and the number of bound proteins becomes low, the probability of crossing the threshold increases because of large high frequency fluctuations in $m$.
Therefore the memory is shorter and precision of the characteristic timescale is lower.

In summary, the regulatory domain architecture set by $N$ and $M$ together with binding kinetics, determine the bound protein dynamics, 
the negative feedback mechanism and the dynamics of the total number of proteins $n_T(t)$.

\section{Discussion}\label{sec_disc}
We studied the stochastic dynamics of a negatively autoregulated gene with multiple binding sites for the repressor,~Fig.~\ref{f_sch}.
We showed that increasing the number of binding sites induces regular bursting of gene products,~Fig.~\ref{f_osc}.
In deterministic systems, explicit time delays or multiple intermediate steps are required together with strong nonlinearity in order to generate biochemical oscillations~\cite{Novak2008, Ferrell2011}.  
The system we study here does not include such intermediate steps or delays yet it produces regular bursting~\cite{Guisoni2016}.
The stochastic dynamics of multiple binding sites occupation introduces an additional effective timescale that triggers regular bursting. 
The vertebrate segmentation clock~\cite{Oates2012} may be a good model system to test our results.
This genetic oscillator acting during embryonic development is thought to be driven by negative autoregulation of \emph{her/Hes} genes, which have several regulatory binding sites for the repressors~\cite{Takebayashi1994, Schroter2012}.
In zebrafish these genetic oscillations appear to be cell autonomous~\cite{Webb2016}.
Although it is thought that autoregulation of \emph{her} genes involves effective delays~\cite{Lewis2003, Schroter2012}, it is possible that the presence of multiple binding sites enhances oscillations~\cite{Lengyel2014}. 
In mouse it has been proposed that collective rhythms are not cell autonomous, but arise as a consequence of intercellular communication from the stabilization and synchronization of bursts~\cite{Masamizu2006}.
The regular bursting induced by multiple binding sites we observe here may underlie these noisy biochemical oscillations and collective rhythms in the mouse segmentation clock.
In summary, our results may establish paradigms for biochemical oscillations that could be tested for example by interfering with binding sites~\cite{Sternberg2015}.

For a fixed number of binding sites we showed that tuning the threshold for autoregulation, noise suppression can be enhanced,~Fig.~\ref{f_noi}.
Noise suppression appears to be strongest when synthesis events are directly driven by fluctuations in binding and unbinding of transcription factors. 
This noise suppression occurs without increasing the product turnover, suggesting a mechanism that could decrease the high cost of regulatory control systems~\cite{Lestas2010}.
Noise can be tamed even when single binding site regulation aggravates fluctuations with respect to unregulated synthesis~\cite{Stekel2008, Gronlund2013, Huang2014}.
Thus, noise suppression can be rescued from this situation without changing the affinity of the molecules for the binding sites.
A model system to test this idea may be \emph{Drosophila}, whose regulatory domains may contain more than 10 binding sites for a single transcription factor~\cite{Berman2002}.

Cellular control systems use negative autoregulation in different contexts either to generate oscillations~\cite{Oates2012, Elowitz2000, Stricker2008}, 
or to control the level of some target molecules by suppressing number fluctuations~\cite{Becskei2000, Dublanche2006, AlonBook}.
Multiple binding sites, occurring in different natural regulatory systems~\cite{Gotea2010, Rydenfelt2014, Takebayashi1994, Schroter2012}, may enhance these functions.
While there may be other mechanisms that perform these feats, as for example cooperativity~\cite{Ferrell2009} and cooperative binding~\cite{Gutierrez2009, Gutierrez2012, Kazemian2013}, 
we argue that adding binding sites to the regulatory region of a gene might be a simpler and more straightforward evolutionary path~\cite{AlonBook}.
Short local duplications of the genome can easily increase the copy number of binding sites for transcription factors~\cite{Thomas2005}.
Microsatellites, which are tandem arrays of multiple copies of a short sequence~\cite{Thomas2005, Li2004}, have a signifficant overlap with transcription factor binding sites~\cite{Iglesias2004, Sinha2005}.
Our results could be useful in the design of synthetic control systems that either generate oscillations~\cite{Elowitz2000, Stricker2008} or tightly control fluctuations around the mean number of molecules~\cite{Becskei2000, Dublanche2006}.

\begin{acknowledgments}
We thank L. Bruno, D. J. J\"org, A. C. Oates and K. Uriu for valuable comments on the manuscript. 
LGM acknowledges support from ANPCyT PICT 2012 1954, PICT 2013 1301 and FOCEM Mercorsur (COF 03/11).
\end{acknowledgments}

\appendix
%
%
\section{Power spectral density and quality factor estimation}\label{sec_hilbert}
Here we outline with more detail the calculation of the power spectral density and the quality factor.
%
%
We run $N_s=100$ stochastic simulations for every set of parameters $\{N,M,r, d_f, k_b, k_u, d_b \}$.
We use the Hilbert transform to obtain a signal without amplitud fluctuations.
We first compute the Hilbert transform of the time series $n_T(t)$ minus the median, $H(n_T(t) -  \text{Med}(n_T))$~\cite{Pikovsky}, 
using the signal processing package \texttt{scipy} of \texttt{python}. %
From the Hilbert transform we extract the analytical phase value and take the cosine of this phase to obtain $h_T(t)$.
We then compute the Fourier transform $\hat{h_T}(f)$ of the time series $h_T(t)$ 
using the Fast Fourirer Transform algorithm implemented in the package \texttt{scipy}.
We obtain its power spectral density 
$ S_{h_T} = | \hat{h_T}(f) | ^2 $~\citep{Stein2000Book}.
We average the power spectral density over the $N_s$ realisations, and fit by least squares the resulting average with a Lorentzian distribution
${\cal L} (x) = {A} {\gamma^2} / (( {\pi\gamma})({(x-f_c)^2+\gamma^2 }) )$,
where $A$ is a scale coefficient, $f_c$ specifies the peak location and $\gamma$ is a dispersion parameter which specifies the half-width at half-maximum.
From the fit we compute the quality factor $Q$ with main frequency $f_0 = f_c$ and half power bandwidth $\Delta f = 2 \gamma$.
\begin{figure}[t]
\centering
\includegraphics[width=\columnwidth]{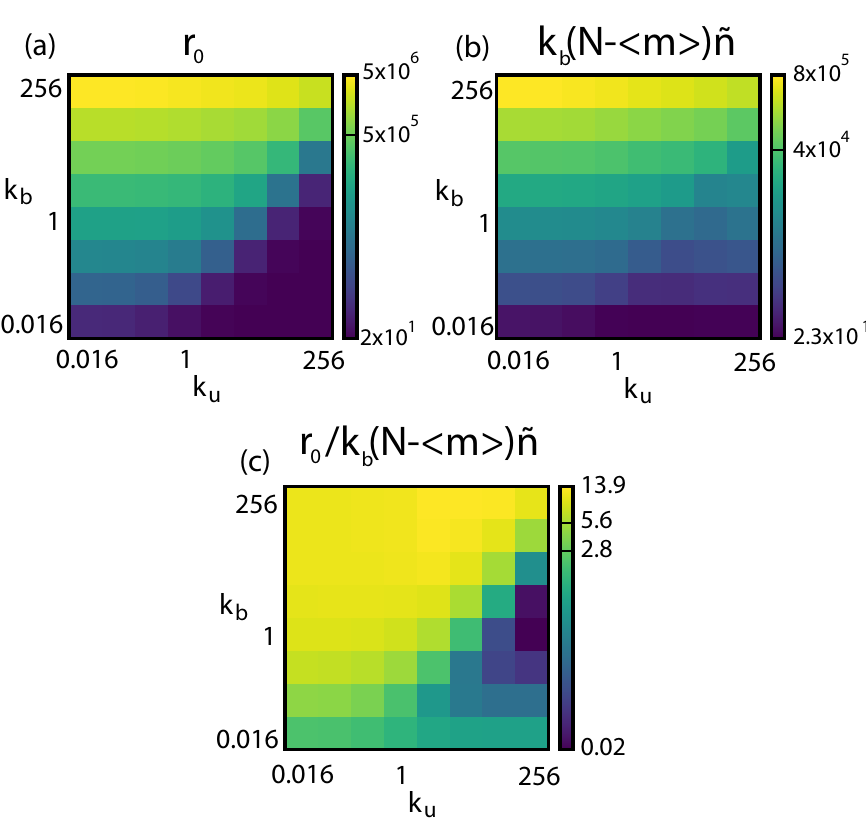}
\caption{
\emph{Large binding rates require large synthesis rates in order to fix the average number of proteins.}
(a) Rate of synthesis $r_0$ for $m\leq M$ for different values of binding $k_b$ and unbinding $k_u$ rates.
(b) Average binding propensity evaluated at burst peaks for different values of binding $k_b$ and unbinding $k_u$ rates.
We introduce $\tilde{n}$, the average of $n$ values at burst peaks.
(c) Propensity ratio for different values of binding $k_b$ and unbinding $k_u$ rates.
Parameters: $N=11$, $M=1$, $d_f=d_b=1$.}
\label{f_r_big}
\end{figure}

\section{Balancing synthesis and binding propensities to fix the average number of molecules}		\label{sec_prop}
To keep the mean number of proteins constant, as we increase the number of binding sites $N$ we also increase the value of the synthesis rate $r_0$. 
Since the binding propensity $a_{\text{bin}}(n,m)=k_b (N-m) {n}$ grows with the number of binding sites $N$,
increasing $N$ increases the mean value of bound proteins $\langle m \rangle$.
Therefore the probability of finding the system open for synthesis $P(m\leq M)$ decreases.
As a consequence there are less synthesis events, so keeping $\langle n_T \rangle$ constant requires a larger synthesis propensity $a_{\text{syn}}(n,m)=r(m)$.
%
%
This balance of increased synthesis rate together with a larger $\langle m \rangle$ maintains the average number of proteins fixed as required.

For fixed $N$, changing the values of the binding and unbinding rates also require a change in the value of $r_0$ since they modify $\langle m \rangle$ and $P(m\leq M)$.
As $k_b $ or $k_u$ vary, the value of $r_0$ spans over several orders of magnitude, but the binding propensity also does, Fig.~\ref{f_r_big}(a,b).
The order of magnitude of the ratio $r_0/a_{\text{bin}}$ remains constant in the region where the noise in the system is higher and the value of $r_0$ is larger, Fig.~\ref{f_r_big}(c).
High synthesis rates are required due to the low probability of synthesising arising from the high binding rate.


%

\end{document}